\documentclass[12pt]{article}
\usepackage{amsmath}
\usepackage{amssymb}
\usepackage{amsthm}
\usepackage[pdftex]{graphicx}
\usepackage{psfrag,epsf}
\usepackage{enumerate}
\usepackage{subcaption}
\usepackage{graphicx}
\usepackage{wrapfig}
\usepackage{indentfirst}
\usepackage{url}
\usepackage{booktabs}
\usepackage{multirow}
\usepackage[colorlinks,citecolor=blue,urlcolor=blue]{hyperref}
\usepackage{longtable}
\usepackage{pdflscape}
\usepackage{authblk}
\usepackage[top=1in, bottom=1in, left=.5in, right=.5in]{geometry}
\usepackage{setspace}
\usepackage{algorithmic}
\usepackage{multirow}
\DeclareUnicodeCharacter{0301}{\'{e}}
 \usepackage{threeparttable}

\allowdisplaybreaks
\usepackage{amssymb}
\usepackage[boxruled,scleft]{algorithm2e}

\RestyleAlgo{boxruled}
\usepackage{mathtools}

\numberwithin{equation}{section}

\usepackage[
backend=biber,
style=numeric,
]{biblatex}

\begin{document}

\title{Inference-based statistical network analysis uncovers star-like brain functional architectures for internalizing psychopathology in children }

\author{Selena Wang\footnote{The correspondence should be addressed to Selena Wang, selena.wang@yale.edu}}
\author{Yunhe Liu}
\author{Wanwan Xu}
\author{Xinyuan Tian}
\author{Yize Zhao}
\affil{Department of Biostatistics, Yale University, New Haven, CT 06511}

\date{}

\maketitle
\doublespace
\abstract{ To improve the statistical power for imaging biomarker detection, we propose a latent variable-based statistical network analysis (LatentSNA) that combines brain functional connectivity with internalizing psychopathology, implementing network science in a generative statistical process to preserve the neurologically meaningful network topology in the adolescents and children population.
The developed inference-focused generative Bayesian framework (1) addresses the lack of power and inflated Type II errors in current analytic approaches when detecting imaging biomarkers, (2) allows unbiased estimation of biomarkers' influence on behavior variants, (3) quantifies the uncertainty and evaluates the likelihood of the estimated biomarker effects against chance and (4) ultimately improves brain-behavior prediction in novel samples and the clinical utilities of neuroimaging findings. We collectively model multi-state functional networks with multivariate internalizing profiles for 5,000 to 7,000 children in the Adolescent Brain Cognitive Development (ABCD) study with sufficiently accurate prediction of both children internalizing traits and functional connectivity, and substantially improved our ability to explain the individual internalizing differences compared with current approaches. We successfully uncover large, coherent  star-like brain functional architectures associated with children's internalizing psychopathology across multiple functional systems and establish them as unique fingerprints for childhood internalization. 

\doublespace

\section*{Introduction}

Childhood and adolescence are critical periods for growth. The development of psychopathology during childhood and adolescence has been linked to significant impairment and mental health issues later in life \parencite{lewinsohn1999natural,beesdo2009anxiety,malone2010adolescent}. Advances in functional magnetic resonance imaging (fMRI) offer new lens to investigate the ontogeny of functional neural systems and uncover novel risk factors of childhood psychopathology. Developing powerful statistical methods linking brain functional network architecture, characterized by functional connectivity with psychopathology is crucial for advancing our understanding of neurodevelopment as well as informing public health policies and promoting adolescent health. This type of research has the potential to effectively intervene and prevent dysfunctions before the start of conventional hallmarks, promote well-being and resilience during developmental stages, key to mitigating the lasting impacts of mental illness.

Despite the clear childhood vulnerability to developing internalizing symptoms, the functional connectivity biomarkers of internalization have remained elusive. The focus on single regions of interest and often isolated connectivity edges to identify functional correlates of the disorder has led to increased attention in regions such as the anterior cingulate, amygdala in intrinsic and actively engaged brains \parencite{ kerestes2014functional,albertina2022internalizing, hall2014fmri,yang2009depressed, kerestes2014functional}. However, concentrating only on localized effects may neglect the multi-faceted nature of developing psychopathology during this critical developmental stage and overlook substantial whole-brain changes
due to widespread restructuring as the brain matures \parencite{krasnegor1997development,paus1999structural,benes1989myelination}. Disjoint and disconnected connectivity biomarkers, may be contributed to low statistical power, are often coupled with inconsistencies of findings across studies with low prediction accuracy and low replicability, impeding the discovery of the exact neurobiological mechanisms of childhood and adolescent psychopathology and effective interventions.

To identify brain connectivity biomarkers, current methods often fit a regression model with vectorized brain connectivity edges as the predictors, and behavioral outcomes as the response. The model is estimated and evaluated in a cross-validated framework to reduce the number of significant connectivity edges from marginal correlation analysis, improve predictability and reduce over-fitting. This class of model is commonly referred as the connectome-based predictive modeling (CPM, \parencite{shen2017using,finn2015functional}). Although CPM and other similar prediction-focused approaches \parencite{mihalik2019abcd} are advantageous due to their computational and conceptual simplicity and have gathered success in predicting cognition \parencite{finn2015functional} and  behavior \parencite{yip2019connectome,lichenstein2021dissociable}, they cannot quantify marker detection uncertainty with controlled Type I errors, ignore the topology and structural dependence of the networked brain and ultimately, can lead to reduced statistical power and misinterpreted results. 

Alternatively, statistical network-based analysis (SNA), a collaboration between network science and statistical theory  \parencite{barabasi2023neuroscience,wilson2021analysis,wang2023establishing,zhao2023genetic,tian2023bayesian,zhao2021bayesian}, has emerged as an advanced analytical framework capable of accommodating the underlying dependent and topological structures of the networked brain while performing various statistical tasks. In contrast to the marginal and univariate edge-wise analyses that achieve biomarker detection via an iterative process, SNA models directly analyze the brain connectivity in its totality as a complex network, thus preserving the neurologically meaningful brain topology information, avoiding reduced statistical power and inflated Type II errors.  Currently, most existing SNA models analyze the brain-behavior relationships as one-sided regression models. These models either analyze the (reduced dimension) brain connectivity as the predictors of a regression with behavior as the response \parencite{zhao2022bayesian,wang2021learning} or analyze connectivity as the response in a matrix-response regression to quantify behavioral covariate effects \parencite{chen2023identifying,shi2016investigating}. However, both types of models lack the ability to capture the mutual variations between behavior profiles and neurodevelopment, i.e., brain development influences children behavior, and abnormal behaviors potentially reinforce the brain abnormality due to brain plasticity \parencite{kolb1998brain,johnston2004clinical,kolb2013brain}. Joint modeling, analyzing the brain-behavior relationship in an unified framework, thus possesses the critical advantage for identifying the functional fingerprints of 
childhood psychopathology as it addresses the high heterogeneity and strong motion artifact in children fMRI data \parencite{chahal2020research,calhoun2016multimodal} and improves prediction accuracy and generalizability of the findings. 

To address this gap, we extend current SNA methods to allow the joint modeling of brain connectivity and behavior in a principled and flexible manner.  In particular, we propose a latent-variable assist statistical network analysis (LatentSNA) framework to simultaneously analyze functional connectivity, collected under different cognitive states, and children internalizing psychopathology, capitalizing on their shared variation. We devise a novel Bayesian Markov chain Monte Carlo (MCMC) algorithm to estimate the model. We apply our method to the large adolescent research population through the Adolescent Brain Cognitive Development (ABCD) study and collectively analyze $5,000$ to $7,000$ brain functional networks under resting state and different cognitive, emotional and behavioral task states with multivariate internalizing measures. Through the analyses, our method comprehensively uncovers and quantifies the uncertainty of the mutual relationships between brain functional alternation and the development of internalizing psychopathology. In addition, using LatentSNA, we demonstrate strong predictive powers in independent samples when predicting either internalizing psychopathology or functional connectivity when only one data component is available in the new sample.


Under LatentSNA, the functional fingerprints of internalization for children are reliably detected to be large, coherent and interconnecting multiple functional systems. 
Using LatentSNA, we substantially improve statistical power, and as a result, we are able to identify a comprehensive set of internalizing biomarkers across multiple functional systems. Our results align with current research suggesting that internalizing psychopathology emerges from intricate interactions and dysregulations within distributed neural networks and systems involved in emotion regulation, cognitive processing, and stress response \parencite{price2012neural,holsboer2000corticosteroid,greicius2007resting}.

While converging evidence suggests that the brain is organized hierarchically with inhomogeneous connectivities and preferential communications among core regions \parencite{nigam2016rich,van2011rich,towlson2013rich}, little is known about the organizational structures of the brain attributable to child and adolescent internalizing psychopathology. 
To the best of our knowledge, this work is among the very first to comprehensively uncover the brain functional network architectures and their organizational principles related to childhood internalization.  Our results show that internalizing psychopathology occurs in children via star-like topological architectures across functional systems \parencite{sawai2012exploring,guimera2002optimal,sawai2014hierarchical} with a few core actors (stars) almost completely connected with each other and many peripheral actors, almost all connected with the star nodes. 
In star-like structures, there are short path distances between the peripheral and the star regions, making the network efficient for communication, robust for transmission failure, and yet prone to congestion.   These star-like architectures suggest that the functional connectivity of a few star regions and cliques can explain individual differences in internalizing psychopathology and serve as the fingerprints to inform the development of internalizing psychopathology and its deterioration. These discoveries support LatentSNA as an important statistical method that can open up neuroscience imaging research with rigorous and powerful analysis.

\section*{Results}

\subsection*{Conceptual framework}

\begin{figure}
    \centering
    \includegraphics[width = 1\textwidth]{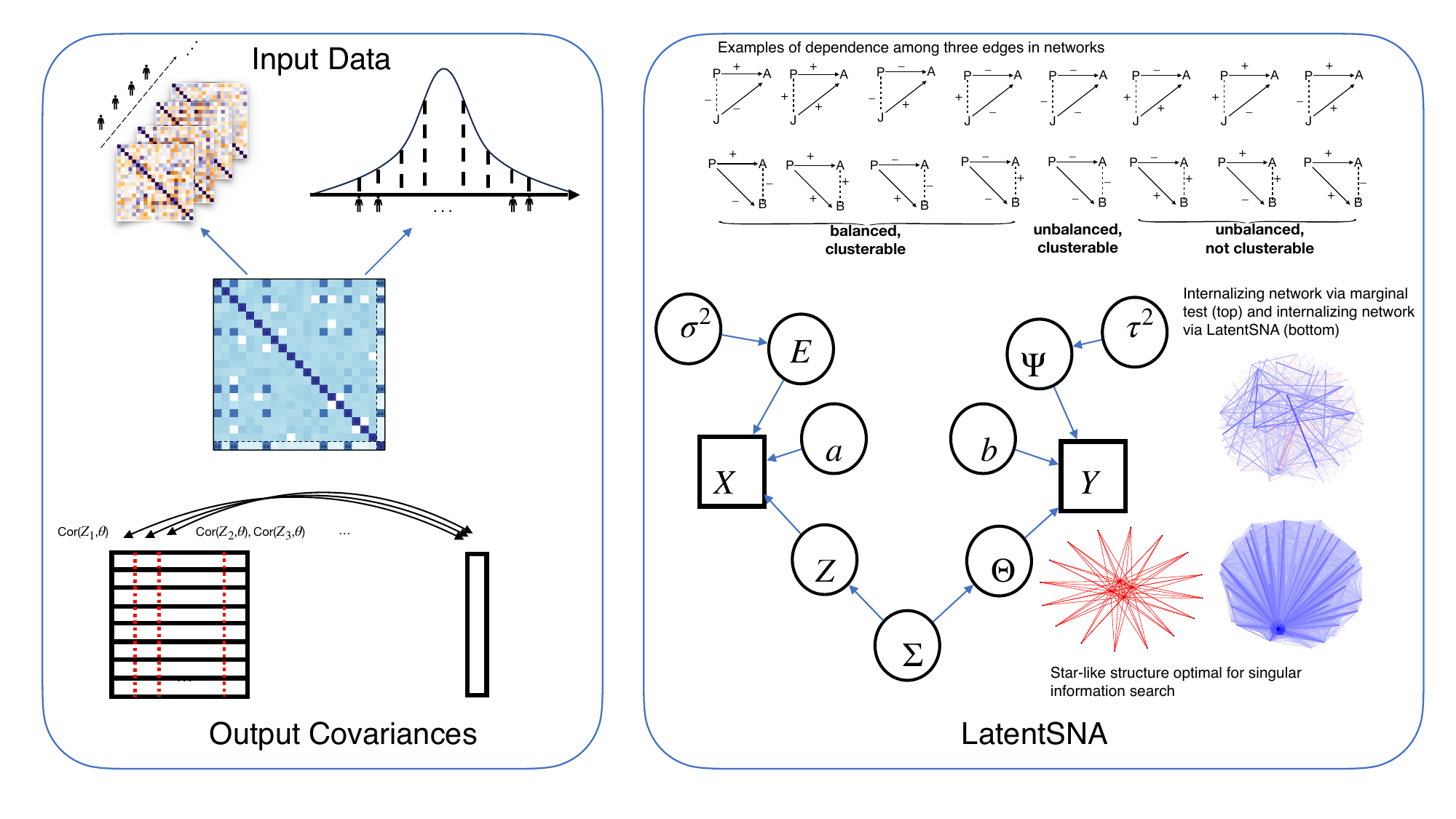}
    \caption{The schematic diagram of LatentSNA. With the input of functional networks and internalizing psychopathology measures, LatentSNA infer region-specific imaging biomarkers of internalizing psychopathology. Circles represent unknown quantities of the model, and squares represent the observed data. Non-informative priors are not included in this diagram.}
    \label{diag}
\end{figure}

We present a schematic diagram showing a few distinctive features of LatentSNA in Figure \ref{diag}. Motivated by the need to improve the power for identifying imaging biomarkers, we propose LatentSNA as a generative statistical network model to characterize the interconnections between brain networks and behavior traits. We propose a \textit{joint} data generation process that allows connectivity differences to inform psychopathological variations, and vice versa---both the brain connectivity and the psychopathology are the targeted modeling interests.  LatentSNA is ideal for detecting complicated and potentially noisy and weak signals hidden in the high-dimensional functional connectivity data. To account for the noise in the connectivity data, we employ latent variables to partial out random noise variations from true signal variations. To reinforce potentially weak signals in the connectivity, we employ a joint modeling framework that allows the true connectivity signals and the true internalizing signals to mutually inform each other, and thus strengthens connectivity signals. To provide uncertainty quantification for biomarker detection, we use a generative modeling framework that focuses on providing robust statistical inference. For a complex phenomenon such as the internalizing psychopathology, what is learned with brain imaging, e.g., fMRI, can be safely assumed to be incomplete, and joint models, on the other hand, offer a more comprehensive and complete picture.

Second, focused on inferring the relationships between functional networks and internalizing psychopathology, LatentSNA is, philosophically, an inference model (also called explanatory models), not a prediction model \parencite{breiman2001statistical,10.1214/10-STS330,shmueli2011predictive}. Inference models are built to describe how potential predictors and explanatory variables explain individual differences in the responses while prediction models ignore this process and focus on the accurate prediction of future responses. Inference models rely on statistical theories such as the central limit theorem and the large sample properties to derive unbiased estimates of the significant effect coefficients with controlled Type I error while prediction models often introduce biases to improve prediction. Inference model is more optimal for detecting imaging biomarkers as it allows us to quantify the uncertainty associated with the identification of imaging biomarkers, not possible with prediction models. With a large enough sample size, our model can, in an unbiased way, identify true mutual relationships between each region's connectivity and internalizing psychopathology with high enough power and controlled Type I error. Meanwhile, prediction-focused methods such as Lasso \parencite{tibshirani1996regression} do not offer unbiased quantification of the relationships, suffers low power and inflated Type II errors.


Third, LatentSNA builds on the statistical network modeling literature and preserves the topological structure of the brain network.  The graphic and dependent structure of the functional connectivity offers unique neurobiological knowledge that informs the learning of the risk network neuromarkers contributing to psychopathology. We make use of the symmetric bilinear interaction effect to capture possible higher-order dependencies, among three or more connectivity edges (see examples of dependencies among three edges in Figure \ref{diag} and \parencite{wasserman1994social,hoff2005bilinear,hoff2008modeling}), because these higher-order dependencies persist in functional connectivity but have been largely ignored in existing edge-wise analyses. While linear additive effects can only capture the variations across the rows and the columns of the network (variation in node degrees), the bilinear interaction effects capture triangular structures of the network and the relatedness among three and more brain regions. High-order dependencies create features of real-world complex networks, such as balance, transitivity and clusterability \parencite{wasserman1994social} that translates to co-activating systems among three or more brain regions, e.g., functional co-activating systems \parencite{shen2010graph}. In particular, the default mode network captures system co-activation, beyond edge-wise, among multiple brain regions. Bilinear effects capture how the distributed patterns of interactions create function and account for the complexity of integrated brain systems not possible with additive effects.  By allowing the model to capture higher-order dependencies beyond edge-wise dependence, we are able to capture features of complex networks such as the star-like structure in functional brain networks (see Figure \ref{diag}).

Finally, LatentSNA offers powerful predictions of both connectivity and behavior variants. We provide a predictive mechanism for internalizing based on connectivity, which simultaneously serves as predictive mechanism for connectivity based on internalizing information. Accurate prediction is achieved by incorporating latent variables to separate signals from noise, by using joint modeling frameworks and allowing information communication between behavior and connectivity during model estimation and by preserving the topology of the brain networks and capturing complex dependence structures not possible with simple linear additive models.


\subsection*{Simulation}

\begin{figure}
    \centering
    \includegraphics[width = .7\textwidth]{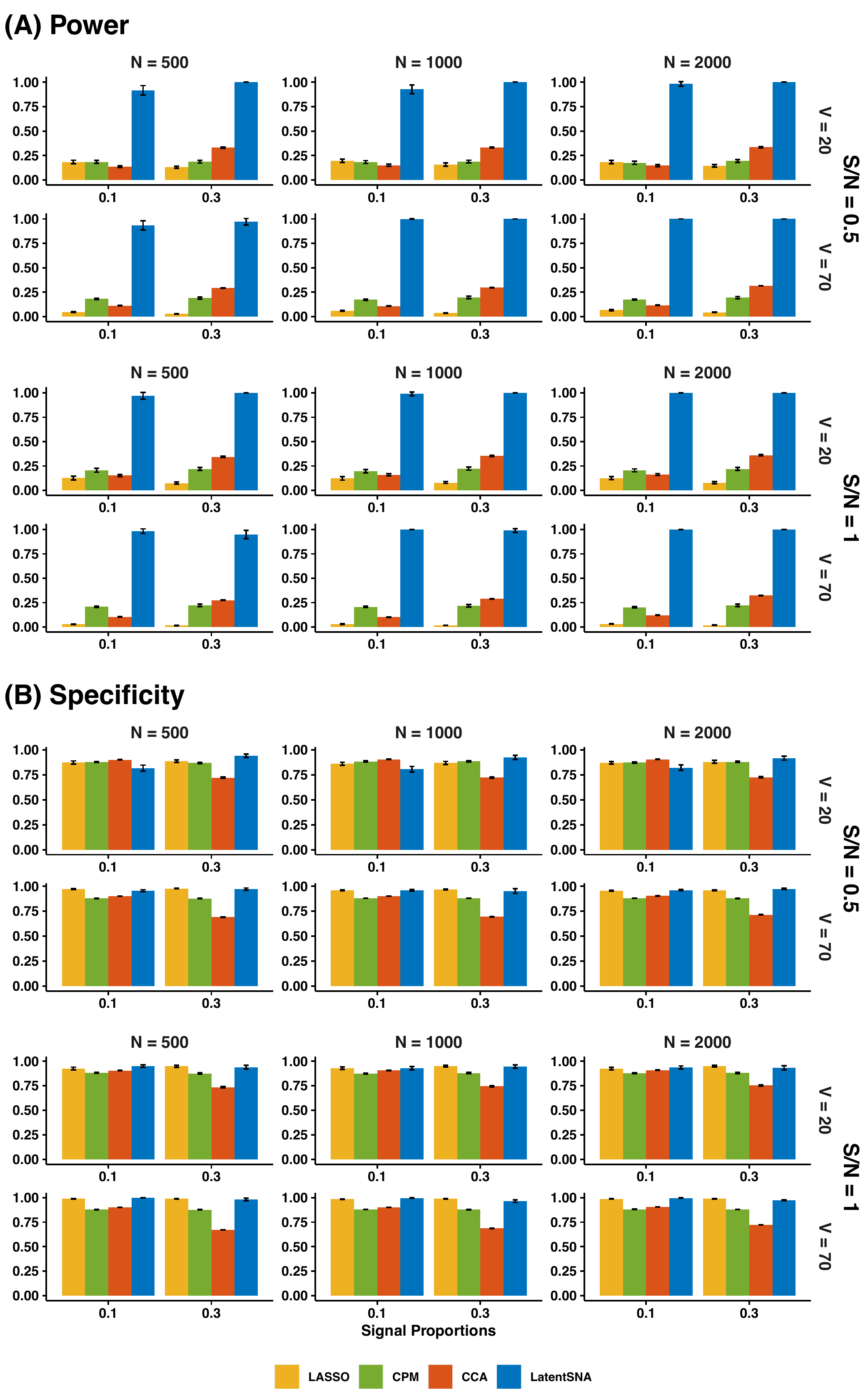}
    \caption{The barplots comparing the power (A) and specificity (B) of LatentSNA with CPM, Lasso and CCA in different data situations. From left to right, the sample size increases from 500, 1000 to 2000.  From top to bottom, we include small ($V=20$) and large ($V=70$) networks, as well as relatively small (0.5) and large (1) signal to noise ratios.  }
    \label{power}
\end{figure}

We compared LatentSNA to CPM, Lasso and the canonical-correlation analysis (CCA), a multivariate method exploring possible dependence between datasets. The comparison was conducted with varying sample sizes, network sizes, signal to noise ratios in the functional connectivity and different levels of relationships (signal proportions) between connectivity and internalizing psychopathology in Figure \ref{power}.  Based on both power and specificity, LatentSNA shows the highest success rate for recovering true relationships and true null relationships making it the most sensitive and accurate method for identifying imaging biomarkers. The relatively low power via CPM reflects the general challenges associated with identifying imaging biomarkers when the fMRI data are noisy and when the relationships between connectivity and psychopathology are sparse. To reduce prediction error, Lasso introduces a penalty term in the loss function, inducing downward bias in the coefficient estimates, and unsurprisingly, reports the lowest power. The high specificity of Lasso is likely the byproduct of the downward bias in the parameter estimation. In comparison, CCA has higher power than Lasso and CPM when there are more relationships between connectivity and internalizing psychopathology. Using CCA, we find the linear combinations of variables on both sides that maximize the dependence between the two, making CCA more powerful when the dependence is strong.  Meanwhile, CCA reports low specificity when the signal proportions are large suggesting that CCA tends to over identify effects with high Type I errors when the relationships between the connectivity and internalizing psychopathology are numerous.

To assess whether the improved power for detection translates to better prediction accuracy of internalizing psychopathology, we report the estimated correlation between the predicted and observed internalizing psychopathology in randomly sampled test data in the supplementary materials. Predicting psychopathology outcomes for a new subject is straightforward within a MCMC sampling scheme described in the Methods section. Using the observed psychopathology $\mathcal{Y}^{(obs)}$ and the current value for the new observation $\mathcal{Y}^{(new)}$, the full conditionals for the latent variable $\theta$ and the model parameters $\boldsymbol{\Psi}$ are unchanged assuming the sampling pattern is ignorable. The full conditional of the new observations $\mathcal{Y}^{(new)}$ is, for any $y_{i} \in \mathcal{Y}^{(new)}$, determined by $\pi (y_{i}|\theta, \boldsymbol{\Psi})$. That is, predicting the psychopathology outcome of a new subject amounts to an additional draw for each missing $y_{i}$ from a normal distribution with probability determined by the model. The results of this process are shown as LatentSNA (THETA) with THETA representing estimated latent internalizing variables.  In addition, we created a LatentSNA (Z) procedure within the LatentSNA framework with Z representing the estimated latent connectivity variables. With LatentSNA (Z), we used the estimated latent connectivity variables, Z, as the predictors to predict the internalizing information in the test data, a predictor-response regression method similar to CPM. Whereas, with LatentSNA (THETA), we directly used the estimated latent internalizing variable for the missing internalizing data  for prediction. Between LatentSNA (Z) and LatentSNA (THETA), the former, similar to CPM, does not directly incorporate training internalizing information for prediction while the latter does.

LatentSNA (THETA) shows the highest prediction accuracy for internalizing psychopathology in most data situations. The prediction accuracy increases as the relationship between functional connectivity and internalizing psychopathology becomes stronger and as the sample size increases. We also present the prediction accuracy of connectivity using LatentSNA in the supplementary materials. Given the duo-predictive capacity of LatentSNA, we are able to rigorously predict each of the testing sample's connectivity given their internalizing information, while in contrast, the Averaging method simply takes the sample average connectivity as a prediction for a new subject's connectivity. In both types of prediction tasks, LatentSNA can accurately predict connectivity networks and internalizing psychopathology in novel samples especially when the connectivity and internalizing psychopathology are strongly related. 
 


\subsection*{Detect functional biomarkers of internalizing psychopathology.}
In this section, we applied LatentSNA to the multivariate internalizing profiles and functional connectivity during the emotional n-back task (EN-back), the Stop Signal task (SST) and the Monetary Incentive Delay (MID) task conditions as well as the resting state (RS) for  5,000 to 7,000 children in the ABCD study.  We aim to uncover functional fingerprints under different cognitive states for childhood internalization, replicate the results, and investigate alternations in the fingerprints between different task and resting conditions.

\paragraph{Accurate prediction of internalizing psychopathology and connectivity in independent samples via LatentSNA.}

To evaluate the accuracy for prediction, we randomly split the dataset into training and test set with 10 random splits, each with a test sample size of $100$ to maintain consistency across task and resting conditions. We then applied the LatentSNA and CPM to predict the
information of the test data using information in the training set  and report the results in Figure \ref{centra}A.  Based on the results, LatentSNA (THETA) shows above $0.9$ median correlation between observed and predicted internalizing information in all four cognitive states, and LatentSNA (Z) shows correlations between $0.6$ and $0.8$. On the other hand, CPM only provides around or below $0.1$ correlations. This strongly supports the advantage of LatentSNA in dissecting reliable predictive information from functional connectivity under each cognitive state. Through those constructed joint learning mechanisms by LatentSNA, we can effectively predict internalizing profiles for new subjects based on the available functional connectivity data.

To assess the prediction accuracy of functional connectivity, we fitted LatentSNA to training  data, calculated the correlation between the observed and the estimated average connectivity and reported the results in Figure \ref{centra}C. 
For 100 test subjects, LatentSNA reports a median correlation of 0.502 for the recovery of the whole brain connectivity, 0.557 for connectivity among the top 10 risk internalizing regions, and 0.707 for connectivity among the top 5 risk internalizing regions. This result shows that LatentSNA provides sufficiently accurate prediction of the connectivity measurements, posing a unique opportunity to uncover brain connectivity for new subjects incorporating their internalizing measures.

\begin{figure}
    \centering
    \includegraphics[width = .75\textwidth]{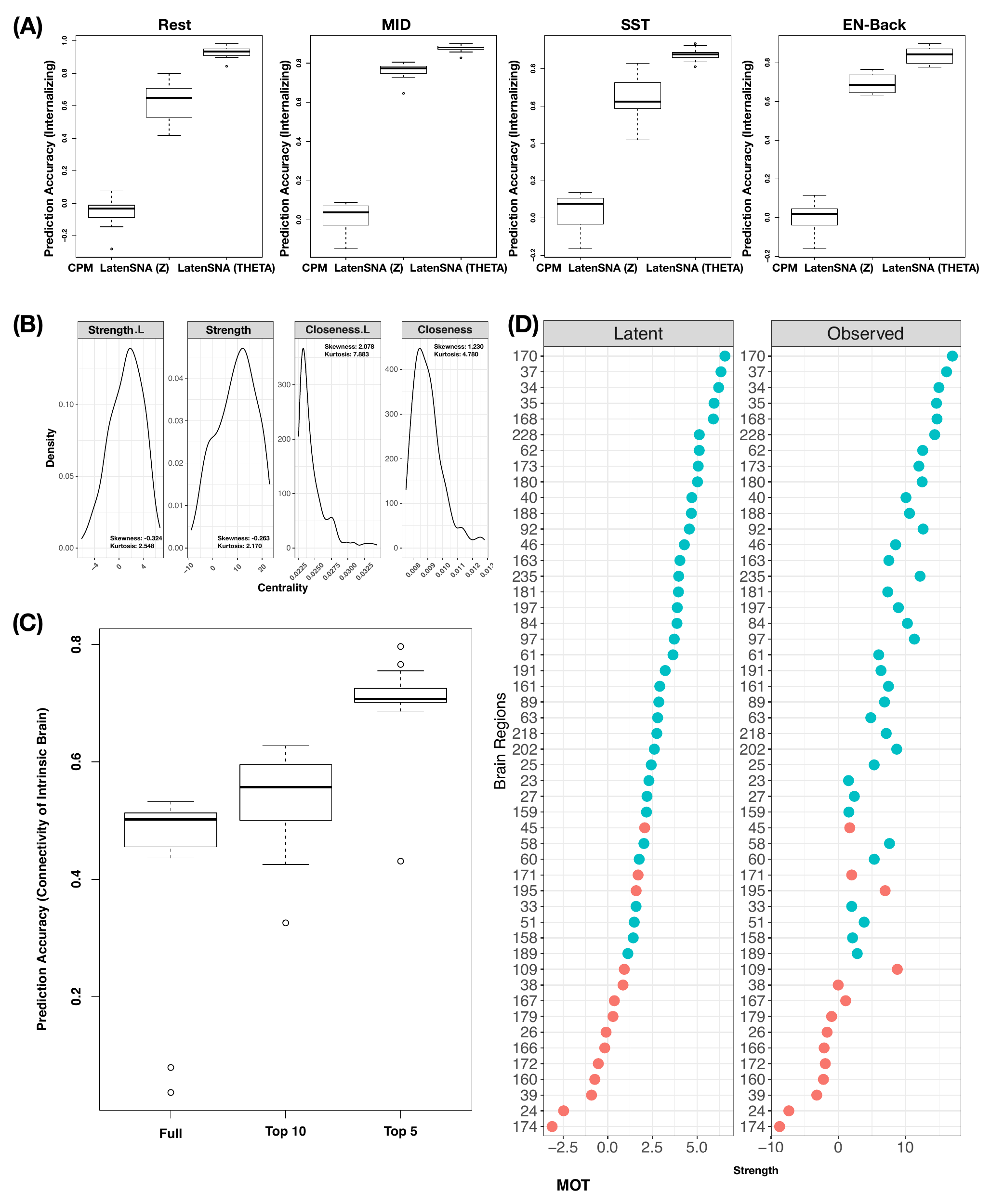}
    \caption{(A) The box plots of the correlations between observed and predicted internalizing psychopathology for each task condition for 100 test subjects  across 10 runs. (B) The densities of the centrality of the brain regions measured by degrees and closeness based on the latent network (left) and the observed network (right) for an average participant during MID condition. (C) The box plots of the correlations between observed and predicted connectivity for 100 test subjects across 10 runs during RS. From left to right, the correlations based on the full networks, networks of top 10 internalizing regions and networks of top 5 internalizing regions are reported. (D) The  strength of each brain region in MOT based on the latent network (left) and the observed network (right) for an average participant during MID condition. Regions identified to play a significant role in explaining individual differences in internalizing behaviors are colored as green, and non-significant regions are colored as red.}
    \label{centra}
\end{figure}

\begin{figure}
    \centering
    \includegraphics[width = 1\textwidth]{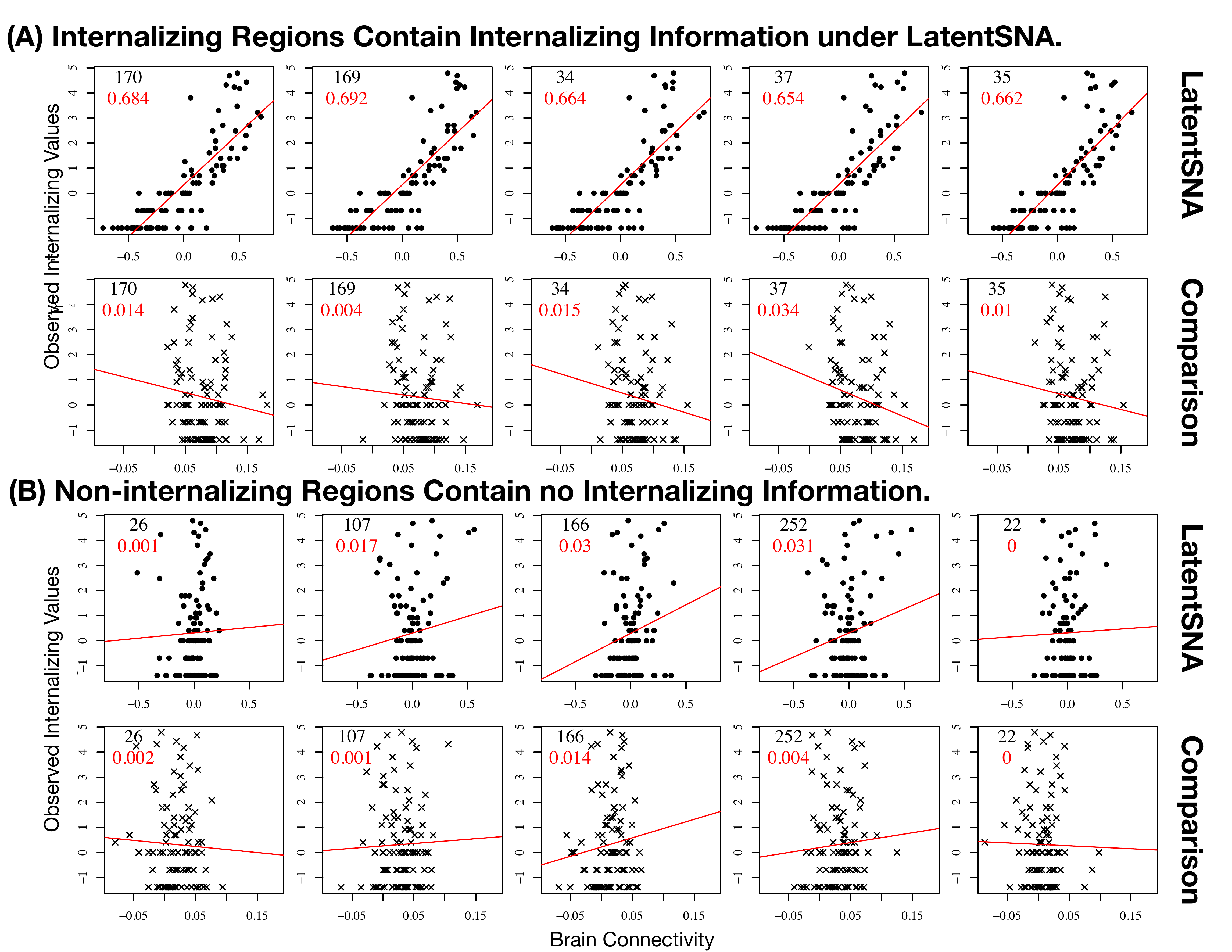}
    \caption{The scatter plots of the observed internalizing values against the connectivity of 5 internalizing regions and 5 non-internalizing regions (region indices are shown at the top left corner in black)  identified by the model. The first and third columns show the scatter plots of the observed internalizing values against the estimated latent connectivity using LatentSNA. The second and forth columns show the scatter plots of the corresponding CPM for a specific brain region, where the predictors are the sums of the significant connectivity edges based on Pearson correlations.}
    \label{scatter}
\end{figure}

 In Figure \ref{scatter}, we compared internalizing regions with non-internalizing regions based on how much they can explain individual differences in internalizing psychopathology in novel samples. We used the sum of the connectivity edges associated with each brain region as the comparison. Estimated via LatentSNA, the latent connectivity variables hold internalizing information and in the LatentSNA-identified internalizing regions, they can explain individual differences in internalizing psychopathology with above $0.65$ correlations, while in non-internalizing regions, they report no association with internalizing psychopathology with around $0$ correlations. This differentiation is absent in the comparison; internalizing regions and non-internalizing regions similarly cannot explain individual differences in internalizing psychopathology. This result demonstrates that LatentSNA differentiates biomarkers from non-biomarkers via the information flow between connectivity and the internalizing information and allows the model-estimated brain connectivity to be informed by internalizing in a data-driven manner.

LatentSNA learns whether a relationship exists between a brain region's functional connectivity and internalizing psychopathology and correctly uses or ignores that relationship depending on whether it exists. In this way, estimated latent connectivity variables contain varying degrees of internalizing information; and a connectivity region contains more internalizing information when it is significantly linked with internalizing psychopathology, and less when it is not linked with internalizing psychopathology.

Within the LatentSNA framework, we modeled internalizing psychopathology as an abstract latent construct (or variable) underlying three observed dimensions of internalizing psychopathology: the anxious-depressed, withdrawn-depressed and somatic complaints dimensions \parencite{stavropoulos2017increased}. These three dimensions represent three conceptually distinct but complimentary manifestations of internalizing psychopathology. Thus, by directly incorporating these dimensions into LatentSNA, we allowed more information to be included than if the internalizing psychopathology is simply modeled as the sum scores of the three dimensions, as is common in the current literature. To assess if the incorporation of the dimensions has improved modeling, we also modeled internalizing as the sum scores using LatentSNA and compare the fit.  We see improvements in predicting internalizing from $0.885$ to $0.825$ for MID, from $0.886$ to $0.893$ for SST, from $0.901$ to $0.846$ for EN-Back and from $0.846$ to $0.791$ for RS when multiple dimensions are directly incorporated. This result suggests that incorporating multiple dimensions of psychopathology is superior to modeling internalizing sum scores.

\paragraph{Large-scale disruptions of multiple functional systems are consistently found with internalizing psychopathology across cognitive conditions.} 
We report the number of significant regions identified in each of the ten functional systems \parencite{shen2010graph} in Figure \ref{task_fig}A, and in Figure \ref{task_fig}B, we show the corresponding $95\%$ credible intervals, the uncertainty quantification obtained from MCMC under the Bayesian inference
of the covariance estimates for each of the 268 brain regions with notable differences between RS and task conditions highlighted in red. Across three task conditions, we found consistent involvement of 131 out of the 268 brain regions and 7 out of 10 functional systems in internalizing psychopathology, supporting internalizing psychopathology as a complex and involving large-scale affective interference of multiple coordinating functional systems. While existing psychopathology literature indicates the involvement of functional systems such as the default mode network, prefrontal cortex, amygdala and other \parencite{price2012neural,kaiser2015distracted,etkin2007functional}, rarely do the studies have large enough power to test the disruption across the whole brain and support the large-scale involvement. Using LatentSNA, we were able to identify and replicate this involvement with other task conditions.

\begin{figure}[!htb]
    \centering
    \includegraphics[width = 1\textwidth]{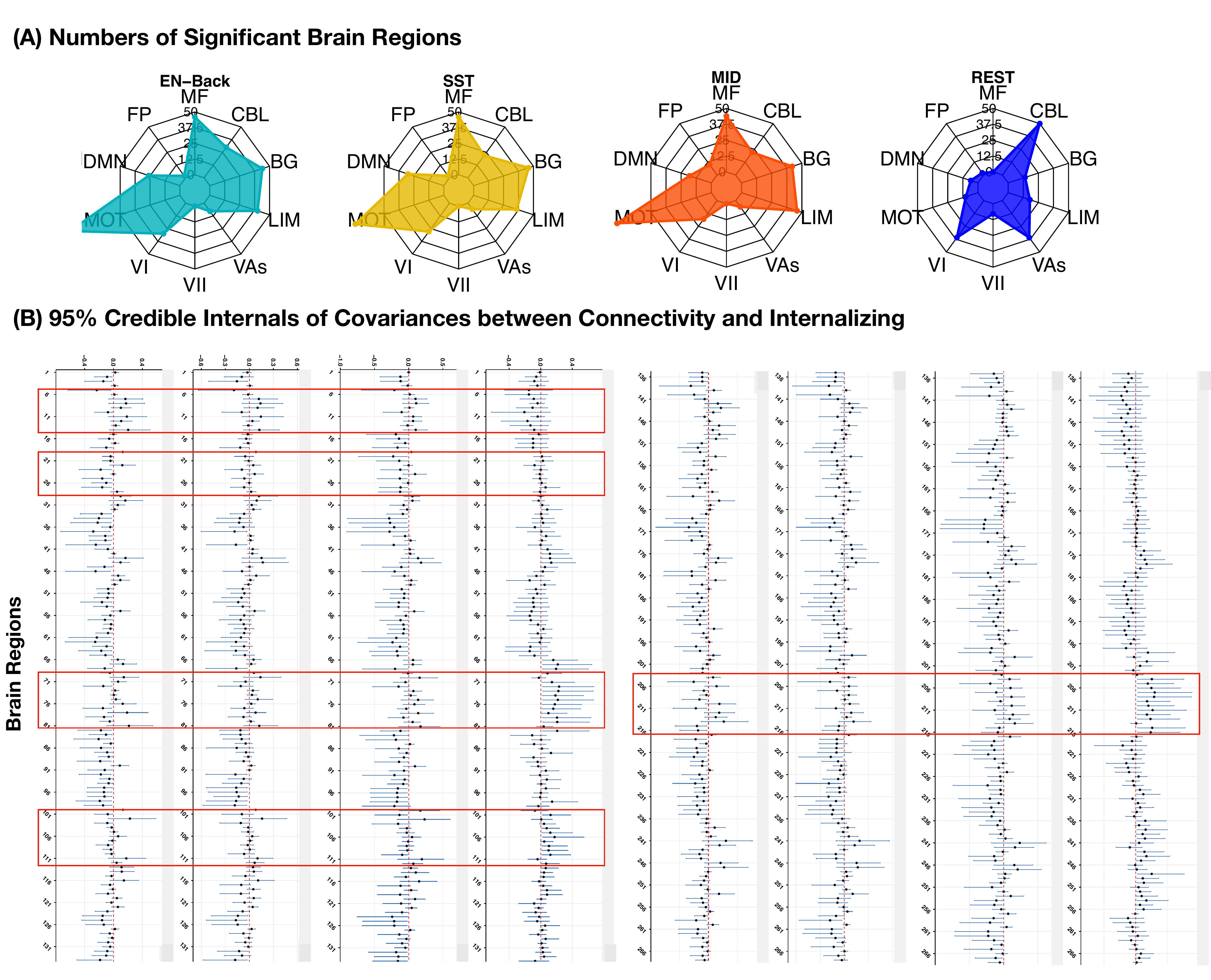}
    \caption{(A) The radar plot showing the number of identified brain regions associated with 10 functional systems for EN-Back, SST, MID and Rest State. (B) The 95$\%$ credible intervals of the covariances between the connectome and internalizing behaviors for each brain region. MF: Medial-Frontal,
FP: Fronto-parietal, DMN: Default Mode , MOT: Motor, VI: Visual I, VII: Visual II, VAs: Visual Association, LIM: Limbic, BG: Basal Ganglia, CBL: Cerebellum.}
    \label{task_fig}
\end{figure}



LatentSNA reveals a shared set of functional architectures attributable to individual variations in internalizing psychopathology when subjects are tasked to perform different emotional and cognitive tasks. This finding corresponds to a recent ABCD study showing similar predictive brain features for various cognitive, personality and mental health scores \parencite{chen2022shared}.  During MID task, the functional connectivity shows the strongest relationship with psychopathology with the highest average covariance estimates (see a comparison of covariance estimates among tasks in the supplementary materials). Figure \ref{task_fig}A shows consistent discrimination of the functional systems and their contributions to developing internalizing psychopathology across tasks. While the motor system, the medial-frontal system, the basal ganglia system, limbic system, default mode network and the visual I systems are consistently found to be implicated in internalizing psychopathology, there is also a consistent lack of implications of the fronto-parietal, visual II and visual association systems. 

The functional architectures of internalizing psychopathology are different for an intrinsic brain vs an active brain. While current literatures support the existence of an intrinsic functional brain during rest with a set of small changes common across tasks \parencite{cole2014intrinsic}, little is known about differences in the functional architectures of internalizing psychopathology under different cognitive states. Our results show evidence for difference in affective interference between RS and task states due to internalizing psychopathology. Different functional connectivity architectures are found to be implicated in internalizing psychopathology between rest and task states.

During RS, three functional systems emerge as the top risk ones to explain individual variations in internalizing psychopathology: cerebellum, visual I and visual association systems. Cerebellum plays an important role in social and emotion processing \parencite{guell2018triple}, and abnormalities are found in cerebellum during rest for subjects with depression \parencite{liu2010decreased} and schizophrenia \parencite{zhuo2018altered}. Our results suggest that during rest, cerebellum is a major functional system contributing to internalizing psychopathology, and its relationship to internalizing is specific to an intrinsic brain, not when the brain is active. Individual differences in the spontaneous functional activities of resting-state visual network, including visual I and visual associations are also related to individual differences in internalizing psychopathology across individuals.

\paragraph{The core-periphery functional network feature is more pronounced with LatentSNA. }LatentSNA differentiates signals from noise in the functional connectivity networks via latent variables. Different from random noise, latent variables capture patterns of meaningful variations in the functional signals across individuals. In LatentSNA, each brain region is allowed to exhibit different levels of variations in the functional signals across individuals and different levels of association with internalizing psychopathology. We captured the true signal variations in reduced dimensions that are much smaller than the dimensions of the network, and we projected these reduced dimensions back to the network dimensions. In this way, we obtained the latent connectivity network capturing the true variations of the functional signals distinct from noise.

Figure \ref{centra}B shows the densities of the node strength and closeness based on the latent network and the observed network for an average participant in the MID condition.  The distributions of the node strength, for both latent network and the observed network, are approximately symmetric based on the d'agostino skewness test \parencite{d1970transformation}. The observed network shows a platykurtic distribution with a significantly negative kurtosis ($p<10^{-5}$, the anscombe-glynn kurtosis test \parencite{anscombe1983distribution}), while the latent network fails to reject the null. A negative kurtosis suggests that the node strength has a flat distribution with thin tails. In comparison, the latent network has more node strength in the tails with more extremely active and extremely dormant regions. Closeness, for both latent network and the observed network, is positively skewed with highly positive kurtosis. Compared with the observed network, the latent network shows larger skewness and kurtosis.  The centrality measures show that the latent network more strongly discriminates core and peripheral regions, reflecting a more pronounced core-periphery differentiation optimal for communication, parallel to those of an efficient information distribution systems \parencite[e.g.,][]{nigam2016rich,stillman2017statistical}.



\begin{figure}
    \centering
   \includegraphics[width = .95\textwidth]{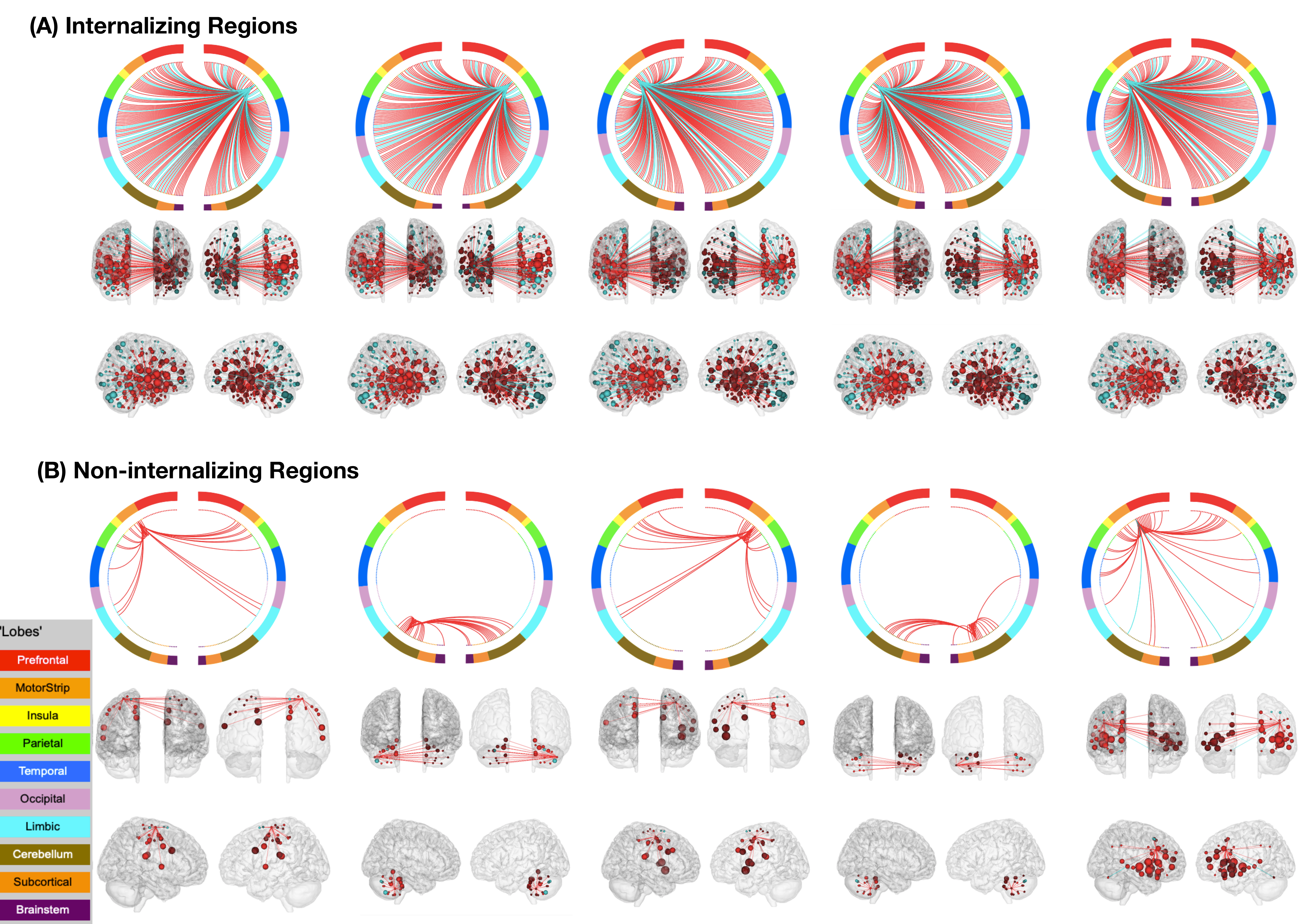}
    \caption{The location and connectivity networks of the top 5 internalizing regions with the strongest relationships with internalizing and non-internalizing regions with no identifiable relationships with internalizing. The circle plots color groups according to anatomical locations, and the 3D brain plots show the front (top left), back (top right), right (bottom left) and left (bottom right) views. }
    \label{conn_fig}
\end{figure}

\paragraph{Functional architectures of internalizing psychopathology are driven by the core actors of the connectivity network.} We report the node strength for regions of the motor system (D) in Figure \ref{centra} and other systems in the supplementary materials. We also show the node strength, closeness and betweenness for all regions in the supplementary materials. The results show that malfunctions associated with internalizing psychopathology are driven by the core actors of the connectivity network. The location and connectivity edges of top five internalizing regions are compared against those of the non-internalizing regions in Figure \ref{conn_fig}. Core regions with high levels of connectivity across the whole brain contribute to the individual differences in internalizing psychopathology. Compared to non-internalizing regions, internalizing regions are the central actors of the functional network with high node strength and high closeness---they are able to transmit a large quantity of information effectively. Development of internalizing psychopathology relies on regions that transmit large quantities of information (high strength) efficiently (high closeness). Low strength and high closeness regions are not internalizing regions---they tend to be the peripheral actors of the network with only localized connectivity edges.

\begin{figure}
    \centering
    \includegraphics[width = .95\textwidth]{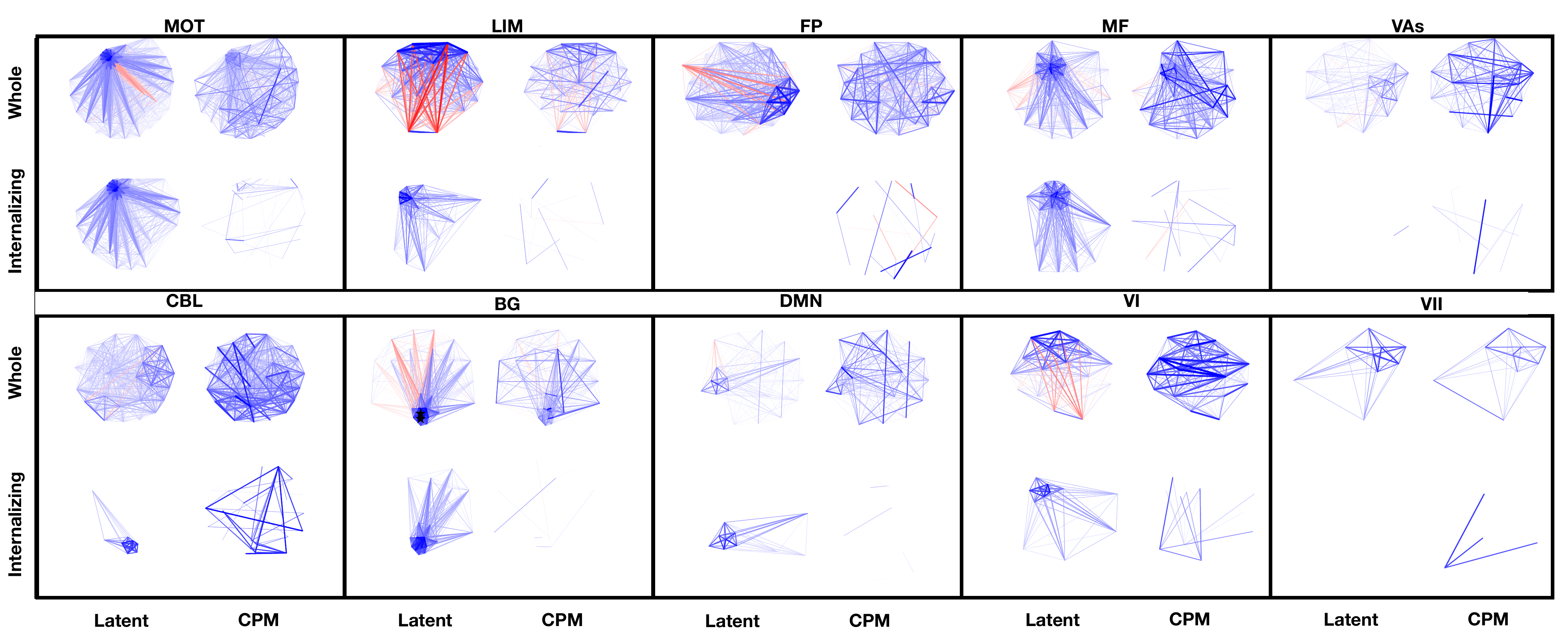}
    \caption{Latent internalizing networks (left) against CPM networks (right) for an average participant in each functional system during MID task. Node positions of the latent networks are then determined using the fruchterman-reingold force-directed graph layout algorithm \parencite{fruchterman1991graph}. The nodes are fixed in the same positions when plotting the internalizing connectivity edges identified via CPM. We also show the corresponding (whole) latent networks, with both significant and non-significant connectivity edges, estimated via LatentSNA, as well as the average observed networks. MF: Medial-Frontal,
FP: Fronto-parietal, DMN: Default Mode , MOT: Motor, VI: Visual I, VII: Visual II, VAs: Visual Association, LIM: Limbic, BG: Basal Ganglia, CBL: Cerebellum. Blue represents the positive connectivity edges, and red represents negative edges.}
    \label{task_fig1}
\end{figure}

\paragraph{Internalizing psychopathology in children are attributable to star-like functional networks.}As the brain is divisible into many coordinating functional systems with distinct connectivity architectures and topology, we report the latent internalizing networks with significant internalizing biomarkers and their connectivity edges in each functional system in Figure \ref{task_fig1}. Star-like structures emerge across functional systems. These star-like structures consist of a few core actors (stars) with many links and many peripheral actors with a few links. The star nodes are almost completely connected with each other forming a central clique, and almost all peripheral nodes are connected with the star nodes. The star-like structure corresponds to the rich club structure often found with brain networks \parencite{nigam2016rich,van2011rich}. In both structures, there are preferential connectivities among core regions. Different from the rich club structure, in the star-like structure, peripheral regions and the central regions are efficiently linked with short path distances, and the peripheral regions are rarely linked among each other with low probability of connections. The star-like configurations contribute to the core-peripheral structure in the latent functional network and the skewness of the centrality distributions. 

The star-like structure is consistent with the current literature on our lack of efficiency when multitasking. The star-like structure is cheap to assemble with a small number of edges and efficient searchability \parencite{guimera2002optimal}. In an ideal one-star network, all peripheral actors are linked with the star, and there is no peripheral-to-peripheral edge. The number of steps to reach an actor in the network is always two, regardless of the network size, making the one-star network the most optimal for communication when only one information search is performed at one time. However, search on the polarized starlike networks quickly becomes expensive when multiple searches occur simultaneously due to congestions at the star nodes.

Our star structure theory provides validity evidence for the current multitasking literature, which supports the idea that the brain is prone to congestions when multiple mental tasks are to be performed \parencite{madore2019multicosts}. The highly connected star regions and their central cliques such as the fronto-parietal control and dorsal attention systems are crucial for completing goal-oriented tasks, but the capacities of these star regions are not limitless. When we multitask, the star regions are likely to be bombarded by competing streams of information with multiple sources of relevant and irrelevant signals, which could lead to congestion. On the other hand, with the star structure theory, the brain is efficiently organized and robust to transmission failure. The star topology reduces the impact of a transmission failure by independently connecting each peripheral region to the star-clique. Peripheral regions communicate with each other via transmission to and from the star. Loss of links between peripheral regions has no impact on the network communication. When there is a failure of transmission between a peripheral region and the star, the peripheral region is isolated, yet the communication on the networked brain is unaffected making it robust to failures.

Due to the efficiency of the brain network communication and its general robustness to transmission failure, degeneration of the functional brain network is damaging when the star regions are compromised. Our results provide evidence for this hypothesis. Both Figures \ref{centra} and \ref{task_fig1} show that internalizing psychopathology in children is attributable to star regions and core cliques of the functional organization. In Figure \ref{task_fig1}, coherent star-like internalizing functional architectures are concentrated in the motor, limbic, medial-frontal, basal ganglia, default node network and visual I functional systems. In comparison, the internalizing functional networks identified through CPM does not exhibit a coherent pattern nor does it follow the central-peripheral differentiation. Our results show that individual differences in the coordinating functional activities of a few star regions can explain substantial individual differences in psychopathology. Thus, star regions' malefactions could have major impacts on development of psychopathology and its further deterioration.


\section*{Discussion}

This work addresses the structures of the functional connectivity networks associated with the development of internalizing psychopathology during childhood and adolescence. Across functional systems, multiple star-like functional architectures are found to contribute to the development of internalizing psychopathology in children, and functional connectivity of the star regions and cliques are found to be associated with the development of internalizing psychopathology. We provide an inference-focused generative framework with the ability to quantify the uncertainty of the identified imaging biomarkers. In addition, we provide a powerful predictive mechanism for internalizing based on connectivity, which simultaneously serves as predictive mechanism for connectivity based on internalizing information. With improved power, we identify large-scale implications of the functional brain associated with the development of internalizing psychopathology during childhood. Our results uniquely contribute to the current literature with the discovery of star-like structures in association with internalizing psychopathology across functional systems. These structures do not work in isolation as disjoint components, instead they closely coordinate and drive emotion regulation and stress response whose disfunction leads to internalizing psychopathology development and deterioration.


Future work should take advantage of the powerful imaging biomarkers detection offered by LatentSNA. Using LatentSNA, researchers can quantify the regional relationships between functional connectivities and internalizing psychopathology in a way that accounts for the dependence structure in the networked brain, and therefore, our detection of the relationships does not suffer lost of power and inflated Type II errors.  LatentSNA is proven to be more powerful than univariate tests and predictive models. Future research should investigate possible biases of the relationships between connectivity and internalizing psychopathology using inference vs predictive models with a formal comparison.
Furthermore, research should also extend LatentSNA to account for longitudinal variations in the functional networks and investigate potential temporal dynamics of fMRI over developmental stages. 

 There are a number of possible extensions of this work. The present specification of LatentSNA only includes region-specific imaging biomarkers of internalizing psychopathology, and future work should consider the group structure among the regions and how regions collectively contribute to internalizing psychopathology---past work has documented the importance of the group structures of the functional brain via functional systems in cognition and disease. 
 
 While we have focused on identifying the relationships between functional connectivity and internalizing psychopathology, our approach can be used to detect other types of biomarkers using positron emission tomography, T1-weighted structural MRI and Diffusion Tensor Imaging, etc. Our approach can also detect imaging biomarkers of other behavioral, cognitive and psychopathology measures. The resulting fits can shed light on a variety of problems, such as the neurodegeneration associated with aging, functional underpinnings of personality and behaviors, and potential groupings of functional organizations together with cognition, behavior and emotion. These and other
possibilities we leave for future work. Finally, beyond neuroscience, the LatentSNA allows the detection of dependence between complex networks and nodal attributes, with applications in many other domains of science. Many complex systems such as social relationships, worldwide webs and transportation grids are impacted by higher-level attributes, and LatentSNA is a statistical technique that can open up many fields with rigorous and powerful analysis.

\section*{Methods}
\paragraph{Participants and Multi-state Functional Connectivity.} We used brain imaging data from the first release of the ABCD study collected from $11,875$ children aged between 9 to 10 years old \parencite{casey2018adolescent}. The blood-oxygen-level-dependent (BOLD) functional activation was recorded for children during resting state (RS) and when they performed three emotional and cognitive tasks. The fMRI data were first preprocessed using BioImage Suite \parencite{joshi2011unified}. The standard preprocessing procedures, such as slice time and motion correction, registration to the MNI template, were described in \textcite{greene2018task} and \textcite{horien2019individual}. The eligible scans had no more than 0.10 mm mean frame-to-frame displacement. Brain images were parceled into 268 regions of interest (ROIs) or nodes using the Shen atlas including
the cortex, subcortex and cerebellum \parencite{shen2013groupwise}. Within each node, the voxel-level time courses were aggregated. Functional connectivity (FC) was constructed for each child in the study during RS and each task state. The FC is constructed by creating a matrix, where each row and column represent all the nodes, and the value of the ($i$, $j$)th entry of the matrix is the Pearson correlation coefficient between the $i$th and $j$th nodes scaled to be normally distributed by a Fisher's z-transformation.

To investigate whether there exists a shared set of neural substrates for internalizing psychopathology across different emotional and cognitive tasks and if the substrates are different from those during rest, we fitted LatentSNA separately to resting functional connectivity and functional connectivity during each task state. We included $7,606$ adolescents with RS functional connectivity capturing intrinsic brain functional activity. We separately investigated the functional connectivity of $4,871$ adolescents who are asked to perform the emotional n-back task (EN-back), $5,096$ adolescents who are asked to perform the Stop Signal task (SST) and $5,298$ adolescents who are asked to perform the Monetary Incentive Delay (MID) task.

\paragraph{Internalizing psychopathology.}Internalizing psychopathology represents a spectrum of conditions characterized by negative emotion including depression, anxiety and phobias \parencite{krueger2006understanding}. In the ABCD study, the internalizing psychopathology is collected via self-reported survey using the Child Behavior Checklist (CBCL, \textcite{stavropoulos2017increased}), which includes 119 items aggregated into 8 empirical
sub-scales. Three sub-scales of CBCL, anxious-depressed (13 items), withdrawn-depressed (8 items) and somatic complaints (11 items) are parts of the internalizing psychopathology. The multivariate representation of the internalizing psychopathology with anxious-depressed, withdrawn-depressed and somatic complaints variables likely outperforms the univariate representation (sum of the three variables) due to the loss of information in the latter. To investigate if this is true, we applied the proposed LatentSNA to both multivariate and univariate representation, and interpret the results based on the model with superior fit, the multivariate internalizing measures.


\paragraph{LatentSNA: latent variables-based statistical network analysis.}Our method makes use of techniques of Bayesian statistical inference, where we propose a generative network model to theorize how the multi-state functional connectivity and internalizing psychopathology intertwine with each other under  random statistical processes with noises. We fitted the functional connectivity networks and accompanying internalizing measures and estimate covariances between the functional connectivity of each brain region with internalizing measures across subjects.

LatentSNA is motivated by the need to improve the power for detecting meaningful biomarkers of psychopathology using noisy imaging connectivity networks. To achieve this aim, we propose LatentSNA with a few distinctive features. First, LatentSNA is a joint model integrating imaging connectivity and behavior or psychopathology variants. Consider a symmetric connectivity tensor, $\mathcal{X} \in \mathbb{R}^{V \times V  \times N}$, where $V$ is the number of nodes for the brain atlas, and $N$ is the number of subjects. Simultaneously, we have information about the behavior or psychopathogy information of the subjects, denoted by the $N \times P$ matrix $\boldsymbol{Y}$, where each row includes the response value for subject $i$ with $p$ psychopathological measurements. 
The proposed LatentSNA is distinct from a network response regression, where the network is the response and the effect of psychopathology on the network is estimated as the regression coefficient of covariates. Similarly, the model differs from a connectivity-based predictive model with psychopathology as the response and the network as the predictor \parencite[e.g.,][]{zhao2022bayesian,wang2021learning}. Instead, we proposed a \textit{joint} data generation process that allows connectivity alternations to inform psychopathological variations, and vice versa---both the brain connectivity and the psychopathology are the targeted modeling interests. 

Second, LatentSNA has roots in statistical network methods and preserves the the topological structure of the network. When modeling the brain connectivity (one of the three components of the model), we made use of the symmetric bilinear interaction effect to capture third-order dependence patterns (transitivity, balance and clusterability) often present in symmetric networks \parencite{hoff2005bilinear,hoff2008modeling}. While additive effects only capture the variations across the rows and the columns of the network (variation in node degrees), the bilinear interaction effects capture triangular structures of the network and the relatedness among multiple brain regions. This is important because these higher-order dependencies exist in functional connectivity. For example, functional systems capture the co-activation of three and more brain regions that creates behavior, cognition and psychopathology. Bilinear effects capture how the distributed patterns of interactions create function and account for the complexity of integrated multimodal brain systems not possible with additive effects. For each subject, we introduced unidimensional region-specific latent variables $z_{u,i}$ to represent connectivity information for subject $i$ and region $u$ and use $z_{u,i} z_{v,i}$ as the driver of connection between brain regions $u$ and $v$ for subject $i$. Each node $u$ is part of a dependent network with strength of connection to node $v$ via the bilinear effect of the two nodes. Specifically, the connectivity between nodes $u$ and $v$, $u < v, u,v = 1,2,.., V$ is modeled by  
    \begin{align}
    &x_{u,v,i} = \boldsymbol{w}_i^T \boldsymbol{\beta} + a_i  +z_{u,i} z_{v,i} + e_{u,v,i}, \qquad e_{u,v,i} \overset{iid}{\sim} N(0, \sigma^2),
\label{connectivity}
    \end{align}
where $a_i$ is the fixed connectivity intercept for subject $i$; $e_{u,v,i}$ is the error term; and $\sigma^2$ is the error variance. We adjusted for $Q$ covariates, e.g., age, gender, denoted by $\boldsymbol{w}_i$ with the first element to be $1$ corresponding to the intercept with their effects on the connectivity matrix characterized by $\boldsymbol{\beta}$. Given that each connectivity value is standardized across persons, node-level additive effects are not necessary. The mean of the connectivity values for each node across persons is zero. In matrix form, we used $\boldsymbol{Z}$ to denote the $N \times V$ matrix of latent variable values, $\boldsymbol{z}_i$ to denote the $V \times 1$ vector of latent variable values for subject $i$, and $\boldsymbol{E}_i$ to denote the $V \times V$ matrix of errors. The approximation of the posterior distributions of the unknown quantities is facilitated by setting  a $\text{MVN} (\boldsymbol{\mu}_{\beta}, \boldsymbol{\Sigma}_{\beta}),  \boldsymbol{\mu}_{\beta} = (0,0,...,0,0)^T, \boldsymbol{\Sigma}_{\beta} = \boldsymbol{I}_{Q} $ prior distribution for $\boldsymbol{\beta}$, a $\text{gamma} (1/2, 1/2)$ prior distribution for $\sigma_e^{-2}$ and a $\text{N} (0, 1)$ prior distribution for $a_i$. The prior for the covariance of the latent network dimensions is described in the joint component.

The third distinguishing feature of LatentSNA is that it focuses on the inference of the relationships between connectivity and psychopathology. For each subject $i$, the probability of pair-wise brain connectivity also depends on the subject's psychopathology $\boldsymbol{y}_i$, and this influence is achieved via joint multivariate normal distribution of the connectivity and psychopathology parameters. Suppose we have $\theta_i$, the unidimensional random latent variable representing the psychopathology information for subject $i$. The connectivity and internalizing information are integrated in the following way: \begin{align}
        &(z_{1,i}, z_{2,i},...,z_{V,i}, \theta_{i})^T  \overset{iid}{\sim} \text{MVN} \left(  \begin{psmallmatrix} \boldsymbol{0}_V \\
  \boldsymbol{0}_D\end{psmallmatrix},
  \boldsymbol{\Sigma}_{V+D}
  \right),  
  \qquad\boldsymbol{\Sigma}= \begin{pmatrix} \boldsymbol{\Lambda}_z &\boldsymbol{\Lambda}_{z\theta}^T\\
\boldsymbol{\Lambda}_{z\theta} &\boldsymbol{\Lambda}_{\theta} \end{pmatrix},
 \label{joint}
\end{align}where $\boldsymbol{\Lambda}_{z\theta}$ is the $V \times D$ matrix modeling the relationship between the connectivity and psychopathology, $D=1$. When there are non-zero elements in the $\boldsymbol{\Lambda}_{z\theta}$ matrix, the connectivity and the attributes regulate and inform each other, which leads to better estimation for both connectivity and psychopathology. Approximation of the posterior distribution of $ \boldsymbol{\Sigma} $ is facilitated by setting a prior distribution of $ \text{Wishart} (\boldsymbol{I}_{V+D}, V+D+2)$. To infer whether the connectivity of a brain region is related to psychopathology, we tested whether the corresponding covariance parameter equals to zero controlling for reflection indeterminancy. We delved deeper into the issue of reflection indeterminancy when discussing estimation. Via the joint distribution, we assume that there is a latent dependence structure between the network and the psychopathology, $ \boldsymbol{\Sigma}_{V+D}$. This dependence structure is region specific with psychopathology having significant links with some brain regions, not others. This dependence structure captures the true (in statistical sense) co-variation between connectivity and psychopathology across individuals, separate from variations due to random noise. If a covariance parameter is significantly different from zero, we can conclude that the associated brain region is significantly linked with psychopathology, and its differences across individuals can explain the individual differences in psychopathology.

Last but not least, using the latent psychopathology variable, LatentSNA allows multivariate modeling of psychopathology with more information to improve its estimation precision  than univariate modeling. In this way, observed psychopathology is generated following a modified version of a psychometric rasch model. The original rasch model \parencite{fischer2012rasch} proposes a data generation process for random test responses in which each test question has a unique difficulty parameter, and each person is ranked based on the number of correct responses. We modified this model in a few ways. The original rasch model does poorly at accommodating data types that are not binary. We included a more flexible linking mechanism for the latent responses and the observed data allowing for both discrete and continuous data distributions. The original rasch model also does not account for covariate effects such gender, race, and to improve, we included a covariate term that allows the probability of responses to vary depending the subject demographics. Most importantly, we introduced a dependence between the measured latent psychopathology variable and connectivity, which allows the measure of psychopathology to be informed by functional connectivity. The degree of dependence is learned via data, and it organically influences how much the psychopathology information are integrated. As the psychopathology component of the joint model, subject $i$'s response on variable $p$ is modeled by
    \begin{align}
        &y_{i,p} = \boldsymbol{h}_i^T \boldsymbol{\gamma} + b_p +\theta_{i} + \epsilon_{i,p}, \qquad \epsilon_{i,p} \overset{iid}{\sim} N(0, \tau^2),  
        \end{align}
where $b_p$ is the fixed intercept for variable $p$. We adjusted for $Q'$ covariates, e.g., age, gender, denoted by $\boldsymbol{h}_i$ with the first element to be $1$ corresponding to the intercept with their effects on the connectivity matrix characterized by $\boldsymbol{\gamma}$. In matrix notation, we used $\boldsymbol{b} $ to denote the $P \times 1$ vector of the intercepts, $\boldsymbol{\Theta}$ to denote the $N \times D$ matrix of latent variables, and $\boldsymbol{\Psi}$ to denote the $N \times P$ matrix of psychopathology errors. As is common in rasch models, the parameters for the question items are fixed, and the person variables are random. Approximation of the posterior distribution of the intercept parameters is facilitated by setting a standard normal prior distribution. We set a prior distribution of $\text{gamma } (1/2, 1/2)$ for $\tau^{-2}$.

\paragraph{Estimation.} Fitting the model involves iterative samples of the full conditional distributions of each parameter defined in the model until we find stable and converged Markov chains to approximate various quantities of the targeted posterior distributions via the Gibbs sampler. Specifically, we iterated the following steps:  \begin{itemize}
        \item simulate $\boldsymbol{\beta}, \boldsymbol{a}$ from their full conditional distributions. 
\item simulate $\sigma^2$ given $\boldsymbol{\beta}, \boldsymbol{a}$, $\tau^{2}$, $\boldsymbol{\gamma}, \boldsymbol{b}$, $\boldsymbol{Z}$, $\boldsymbol{\Theta}$,   $\boldsymbol{\Sigma}$, $\boldsymbol{X}, \boldsymbol{Y}$. 
\item simulate $\boldsymbol{\gamma}, \boldsymbol{b}$ from their full conditional distributions. 
\item simulate $\tau^{2}$ given $\boldsymbol{\beta}, \boldsymbol{a}$, $\sigma^2$, $\boldsymbol{\gamma}, \boldsymbol{b}$, $\boldsymbol{Z}$, $\boldsymbol{\Theta}$,   $\boldsymbol{\Sigma}$, $\boldsymbol{X}, \boldsymbol{Y}$. 
\item simulate $\{ \boldsymbol{Z} \text{ and } \boldsymbol{\Theta}\}$ from their full conditional distributions. 
\item simulate $\boldsymbol{\Sigma}$ from its full conditional distribution. 
\end{itemize} To allow the information in connectivity and psychopathology to flow between each other and mutually inform parameter estimation, we sampled $\{ \boldsymbol{Z} \text{ and } \boldsymbol{\Theta}\}$ from their joint full conditional distribution given both the connectivity and the  psychopathology. For subject $i$, the joint full conditional distribution of $\boldsymbol{z}_{i}$ and $\theta_i$ is the product of the three parts (connectivity, psychopathology and joint):  \begin{align}
p \left( \begin{psmallmatrix}\boldsymbol{z}_{i} \\ \theta_i\end{psmallmatrix} |  \boldsymbol{t}_i, \boldsymbol{\tilde{f}}_{u,i} , \boldsymbol{\Sigma}, \sigma^2_{\epsilon}\right) \propto& p ( \boldsymbol{t}_i | \theta_i, \sigma^2_{\epsilon}) p( \boldsymbol{\tilde{f}}_{u,i} | z_{u,i} ) p\left(\begin{psmallmatrix} \boldsymbol{z}_i \\\theta_i\end{psmallmatrix} | \boldsymbol{\Sigma} \right) \nonumber \\
 \propto& \exp \left( -\frac{1}{2} \sigma^{-2}_{\epsilon}  \sum_{p=1}^P  (t_{i,p} - \theta_i )^2      \right)  \exp \left( -\frac{1}{2}   \sum_{v=1, v \neq u}^V ( \tilde{f}_{u,v,i} - c z_{u,i}^T z_{v,i}  )^2 
\right)\nonumber \\
& \exp \left(  -\frac{1}{2}  \begin{psmallmatrix}\boldsymbol{z}_{i} \\ \theta_i\end{psmallmatrix}^T \boldsymbol{\Sigma}^{-1}   \begin{psmallmatrix}\boldsymbol{z}_{i} \\ \theta_i\end{psmallmatrix} \right)\nonumber,
\end{align}where $\boldsymbol{T} = \boldsymbol{Y} - \boldsymbol{1} \boldsymbol{b}^T -  \boldsymbol{H}\boldsymbol{\gamma}\boldsymbol{1}_P^T $, and $\boldsymbol{F}_{i}$ is $ \boldsymbol{X}_i- a_i - \boldsymbol{w}_i \boldsymbol{\beta} = \boldsymbol{z}_i \boldsymbol{z}_i^T + \boldsymbol{E}_i.$ We can transform $\boldsymbol{F}_i$ in such a way that the transformed error term is a standard normal distribution using $
\boldsymbol{\tilde{F}}_i = c \boldsymbol{F}_i $, 
where $
c= \sigma_e^{-1}$. Therefore,  $
\boldsymbol{\tilde{F}}_i = c \boldsymbol{z}_i \boldsymbol{z}_i^T + \boldsymbol{\tilde{E}}_i, 
$ where $\tilde{e}_{u,v,i}$ follows a standard normal distribution. The joint part of the distribution  $p\left(\begin{psmallmatrix}z_{u,i} \\\theta_i\end{psmallmatrix} | \boldsymbol{\Sigma}' \right)$  can be written as $
\exp (-\frac{1}{2}  (z_{u,i} Q'_{z} z_{u,i} + z_{u,i} Q'_{\theta z} \theta_i +\theta_i Q'_{z \theta}z_{u,i}+ \theta_i^T Q'_{\theta} \theta_i 
  ))$, where $
\boldsymbol{\Sigma}^{-1} = \begin{pmatrix} Q_z &Q_{\theta z}\\ 
Q_{z \theta} & Q_{\theta}\end{pmatrix}
$---each component is a function of $\Lambda$s---, and $\boldsymbol{\Sigma}' $ is part of $\boldsymbol{\Sigma} $ only involving the specific brain region.  
Extracting relevant terms from $p \left( \begin{psmallmatrix}\boldsymbol{z}_{i} \\ \theta_i\end{psmallmatrix} |  \boldsymbol{t}_i, \boldsymbol{\tilde{f}}_{u,i}, \boldsymbol{\Sigma}, \sigma^2_{\epsilon}\right)$, we can see that the full conditional distribution of $z_{u,i}$ is
\begin{align}
&p \left( z_{u,i} | \boldsymbol{\tilde{f}}_{u,i}, \boldsymbol{\Sigma},\theta_i \right) \nonumber \\
  \propto & \exp 
\left(  -\frac{1}{2} z_{u,i}   (  \sum_{v=1, v \neq u}^V c^2 z_{v,i} z_{v,i} + Q_z' ) z_{u,i}
 + z_{u,i}^T (\sum_{v=1, v \neq u}^V c \tilde{f}_{u,v,i}z_{v,i} -\frac{1}{2}  
 Q'_{\theta z}\theta_i -\frac{1}{2} Q_{z y}^{'T}\theta_i) 
 \right),
\end{align} a multivariate normal distribution, with variance $( \sum_{v=1, v \neq u}^V c^2 z_{v,i} z_{v,i} + Q'_z )^{-1}$ and mean $ ( \sum_{v=1, v \neq u}^V c^2 z_{v,i} z_{v,i} + Q'_z )^{-1}
(\sum_{v=1, v \neq u}^V c \tilde{f}_{u,v,i}z_{v,i} -\frac{1}{2}  
 Q'_{\theta z} \theta_i -\frac{1}{2} Q_{z y}^{'T} \theta_i) $. The latent variable value for psychopathology is informed by brain connectivity and should be sampled from
\begin{align}
&p \left(\theta_i | \boldsymbol{t}_i, \boldsymbol{\Sigma}, z_{u,i},  \boldsymbol{A}, \sigma^2_{\epsilon}  \right) \nonumber \\
  \propto & \exp 
\left(  -\frac{1}{2}\theta_i^T   ( \sigma^{-2}_{\epsilon} \sum_{p=1}^P \boldsymbol{\alpha}_p \boldsymbol{\alpha}_p^T + Q_{\theta} )\theta_i
 +\theta_i^T (\sum_{p=1}^P  \sigma^{-2}_{\epsilon}  t_{i,p}\boldsymbol{\alpha}_p -\frac{1}{2}  
 Q_{\theta z}^T \boldsymbol{z}_i -\frac{1}{2} Q_{z \theta}\boldsymbol{z}_{i}) 
 \right),
\end{align} a multivariate normal distribution, with variance $(\sum_{p=1}^P  \sigma^{-2}_{\epsilon} \boldsymbol{\alpha}_p \boldsymbol{\alpha}_p^T + Q_{\theta} )^{-1}$ and mean $ (\sum_{p=1}^P  \sigma^{-2}_{\epsilon} \boldsymbol{\alpha}_p \boldsymbol{\alpha}_p^T + Q_{\theta} )^{-1}
(\sum_{p=1}^P t_{i,p}  \sigma^{-2}_{\epsilon} \boldsymbol{\alpha}_p -\frac{1}{2}  
 Q_{\theta z}^T \boldsymbol{z}_{i} -\frac{1}{2} Q_{z \theta}\boldsymbol{z}_{i})  $.
Crucially, we sampled the covariance matrix  $\boldsymbol{\Sigma}$ from  a inverse-Wishart $(\boldsymbol{I}_{V+D} + \boldsymbol{F'}^T \boldsymbol{F'}, N+V+D+2)$ with $\boldsymbol{F'}$ as a $N \times (V+1)$ matrix with $i$th row as $(\boldsymbol{z}_{i}^T, \theta_{i}^T)$.

The introduction of the bilinear effect, $z_{u,i} z_{v,i}$ induces partial reflection indeterminacy.  For each set of latent variable values,  $\hat{z}_{u,i}$ and $\hat{z}_{v,i}$, the positions given by $-\hat{z}_{u,i}$ and $-\hat{z}_{v,i}$ give the same set of product and consequently the same likelihood. During the MCMC chain, the sign of  $z_{u,i}, u=1$ can change while maintaining the same connectivity value. Crucially, the connectivity latent variables are also related to psychopathology, whether $z_{u,i}$ is estimated as $\hat{z}_{u,i}$ or $-\hat{z}_{u,i}$ has consequences on the correlation between  $z_{u,i}$ and $\theta_i$. Put in a different way, $z_{u,i}$ is softly identified as the signs of $z_{u,i}$ need to satisfy the correlation between $z_{u,i}$ and $\theta_i$. To estimate such a model, we assume that after a sufficient burn in period, the signs of $z_{u,i}$ have reached a sufficiently optimal point, where its correlation with $\theta_i$ has researched a stabilized estimate resembling the true correlation. After this burn in period, we fix the signs of $z_{u,i}$ to the same as target, i.e. target $=$ estimated $z_{u,i}$ from the first iteration after burn in. Therefore, there is no reflection indeterminancy issue after burn in.

\paragraph{Simulation.} The synthetic data are generated following the model formulation. The connectivity and the psychopathology latent variables are generated from the multivariate normal distribution with mean zero and the pre-defined covariance matrix with unit variances. When signal proportion equals to $0.1$ ($0.3$), we randomly assigned $10\%$ ($30\%$) of the covariance parameters between connectivity and psychopathology to be $0.9$, as well as the corresponding values in the latent connectivity covariance matrix to ensure the positive definiteness of $\boldsymbol{\Sigma}$. We randomly sampled the errors for the connectivity from the normal distribution with mean 0 and variance defined by the signal to noise ratio. The errors for the psychopathology are sampled from the normal distribution with the mean 0 and variance 0.5. For simplicity, the number of latent variables in both connectivity component and psychopathology component are assigned as 1 in all generated data. We generated 100 datasets for each data situation (see simulation section in main). We compared LatentSNA against connectome-based predictive model (CPM), Lasso and canonical-correlation analysis (CCA). For Lasso, we fitted the model to the training set using the glmnet package \parencite{friedman2009glmnet}. We selected significant edges based on minimizing mean squared error with $10$ fold cross-validation. For CCA, we fited the model to the training set using the CCA package \parencite{gonzalez2008cca}, and regions with strong loadings are considered to be related to psychopathology. The cut-off thresholds are determined by the true signal proportions. For example, when the true signal proportion equals to $0.1$, we considered top $10 \%$ of regions with highest absolute loadings to be significantly linked with psychopathology.


\paragraph{Implementation and Evaluation.} The estimation algorithm for this paper was implemented in the open sourced programming language for statistical computing and graphics, R. The code will be made available upon acceptance. For each task condition, we performed posterior inference based on the MCMC algorithm under random initials.  
No obvious non-convergence issues were found via trace plots.  For each task condition, we compared the model fit of the multivariate internalizing psychopathology with that of the univariate internalizing psychopathology. The univariate internalizing psychopathology is the sum of the three internalizing variables mentioned before.

LatentSNA can be used to predict future connectivity and behavior variants. To estimate the probability of a connectivity link and the probability of behavior observations, we used the estimated posterior means of model parameters and follow the proposed model equations. To evaluate the prediction, we randomly split the dataset into training and test sets. We then predicted the
information of test set using information in the training set.

While it is not the most optimal to predict entirely missing connectivity information for a subject using the averaging method 
given that there is no information available about this subject in the model, we can use the LatentSNA and rely on the subject's related behavior information. The missing connectivity predictions are useful when individuals’ imaging information is entirely missing or when we need to predict potential future changes in the brain network following developmental changes in cognition. In this scenario, we used the latent connectivity estimated with the missing individuals’ behavior information, other individuals’ behavior, and other individuals’ brain connectivity. We can also predict individuals’ missing behavior information based on their brain connectivity and other individuals'
behaviors and brain connectivity. The missing behavior predictions are useful when individual’s behavior information is
entirely missing or when we need to predict potential future changes in the behavior following changes in connectivity. For CPM, we fitted the model to the training set using the NetworkToolbox R package \parencite{christensen2018networktoolbox} using $10$-fold cross-validation. Each of these procedures is repeated for $10$ times to show variability. 

The node strength, an extension of degree in weighted networks, is the sum of the edge weights associated with each node \parencite{newman2001scientific,opsahl2010node}. Closeness reflects how quickly one node can reach others. We calculated closeness in the weighted graphs using the igraph R package \parencite{freeman1979centrality,csardi2006igraph}, and a uniform magnitude equaling the largest negative edge is added to all edges to ensure that all weights are positive. Among the shortest paths in a network that pass through intermediate nodes, betweenness reflects how many times a node is present in those paths and demonstrates the extent to which a node is part of connections among other nodes \parencite{freeman1977set}. We calculated the betweenness of the connectivity networks with positive weights defined as before using igraph R package \parencite{brandes2001faster,freeman1979centrality}. High betweenness reflects power as it positions the region with an important bridging role allowing the neighboring regions to connect \parencite{burt2004structural}, an investment into the communication between distant clusters.

\newpage

\printbibliography

\newpage

\end{document}